\documentclass{revtex4}

\usepackage{amsmath,amssymb}
\usepackage{graphicx}
\usepackage{textcomp}
\usepackage{units}

\begin{document}

\title{Tunability of Critical Casimir Interactions by Boundary Conditions}

\author{Ursula Nellen}
\affiliation{2. Physikalisches Institut, Universit\"at Stuttgart,
Pfaffenwaldring 57, 70550 Stuttgart, Germany}

\author{Laurent Helden}
\affiliation{2. Physikalisches Institut, Universit\"at Stuttgart,
Pfaffenwaldring 57, 70550 Stuttgart, Germany}

\author{Clemens Bechinger}
\affiliation{2. Physikalisches Institut, Universit\"at Stuttgart,
Pfaffenwaldring 57, 70550 Stuttgart, Germany}
\affiliation{Max-Planck-Institut f\"ur Metallforschung,
Heisenbergstr. 3, 70563 Stuttgart, Germany}

\date{\today}

\begin{abstract}
We experimentally demonstrate that critical Casimir forces in
colloidal systems can be continuously tuned by the choice of
boundary conditions. The interaction potential of a colloidal
particle in a mixture of water and 2,6-lutidine has been measured
above a substrate with a gradient in its preferential adsorption
properties for the mixture's components. We find that the
interaction potentials at constant temperature but different
positions relative to the gradient continuously change from
attraction to repulsion. This demonstrates that critical Casimir
forces respond not only to minute temperature changes but also to
small changes in the surface properties.
\end{abstract}


\maketitle

In 1978 Fisher and de Gennes pointed out that if two objects are
immersed in a fluid close to its critical point, long-ranged forces
due to confined critical fluctuations act between their surfaces
\cite{fis78}. Such critical Casimir forces arise due to the
confinement of fluctuations in the order parameter of the fluid
between the objects. In the case of e.g. a classical binary liquid
mixture close to its demixing point, the order parameter corresponds
to the concentration difference between the two components of the
mixture. The strength and range of critical Casimir forces is set by
the fluid's bulk correlation length $\xi$ which diverges upon
approaching the critical temperature $T_{C}$. Therefore, close to
$T_{C}$, the interaction strongly depends on the temperature as has
been recently confirmed in several experiments
\cite{gam09b,her08,raf07,gan06,fuk05,gar02,gar99,muk99}.

In addition to their temperature dependence, critical Casimir forces
are very sensitive to the boundary conditions (BC) which are
determined by the adsorption preferences of the mixture's components
at the confining surfaces: not only the magnitude, but even the sign
of critical Casimir interactions can be altered by corresponding
symmetric or asymmetric BC. So far, theoretical studies largely
concentrated on BC, where one species of molecules in the binary
liquid mixture forms a saturated monolayer at the confining surfaces
\cite{han98,kre97}. Depending on whether both surfaces strongly
adsorb the same $(- -)$ or different species $(- +)$, this results
in attractive or repulsive forces which have been recently observed
in several experiments
\cite{tro09,soy08,raf07,gan06,fuk05,gar02,gar99,muk99}.

In this Letter we report the first critical Casimir measurements for
continuously tunable boundary conditions. This has been achieved by
measuring the interaction energy of a single colloidal particle
suspended in a critical water-2,6-lutidine mixture above a solid
surface with a gradient in its adsorption preference for the two
liquid components. Upon lateral displacement of the particle
relative to the substrate we find a smooth transition from
attractive to repulsive critical Casimir forces. The observed
scaling functions are found to lie between that of the limiting
cases of $(- -)$ and $(- +)$ BC.

\begin{figure}
\includegraphics[width=8.5cm]{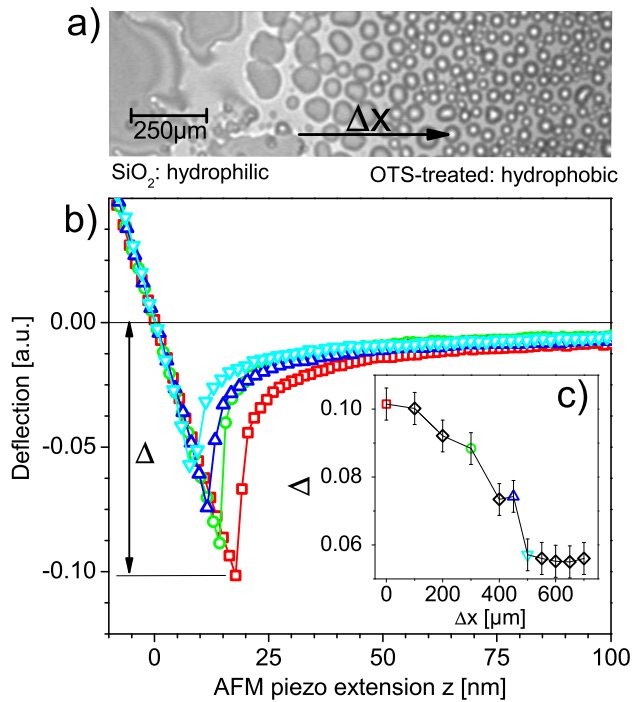}
\caption{(Color online) (a) Water droplets on a silica substrate
with a gradient in its wetting properties below the dew point
(breath figure). Small and large droplets indicate hydrophobic and
hydrophilic regions. (b) AFM force-distance curves obtained with a
hydrophilic cantilever at different positions $\Delta x$ on the
surface. Curves were arranged such that the regions of constant
compliance overlap at zero deflection signal. (c) Attraction
strength $\Delta$ vs. $\Delta x$ along the chemical gradient.
Symbols correspond to those used in (b).} \label{fig:1}
\end{figure}

Surfaces with a spatial variation of adsorption preference for
lutidine and water molecules were fabricated by immersing
hydrophilic silica substrates into a mixture (1300:1) of hexane and
octadecyltrichlorosilane (OTS). After about 30 minutes, a monolayer
of OTS molecules binds to the surface and thus renders it
hydrophobic \cite{ulm96}. Measurements of the contact angle confirm
that this treatment alters the adsorption preference from that of
water to lutidine. We obtained samples with a smooth lateral
gradient regarding the OTS coverage by partially shielding the
substrate with a thin metal blade and exposing it to an
oxygen-nitrogen plasma, so that OTS molecules are fractionally
removed. This gradient can be visualized by cooling the sample below
the dew point. The corresponding breath figure (Fig.~\ref{fig:1}a)
shows small droplets with a large contact angle on the hydrophobic
side where the sample was fully covered by the blade (right) and
much larger droplets with small contact angles (left) where the OTS
molecules were removed and the sample is hydrophilic. The gradient
in the surface properties was further characterized by
force-distance curves obtained with an atomic force microscope (AFM)
under ambient conditions and a freshly plasma-cleaned hydrophilic
tip. Upon approaching the surface, at small distances the tip is
suddenly attracted towards the surface due to van der Waals and
capillary forces (Fig.~\ref{fig:1}b). The strong contribution of
capillary forces is supported by the fact that both strength
$\Delta$ and range of the attraction decrease towards the
hydrophobic side, i.e. with increasing $\Delta x$
(Fig.~\ref{fig:1}c). Similar to the breath figures, the
AFM-measurements show that the above method yields substrates with
smooth chemical gradients which laterally extend over a distance of
several hundreds of microns.

We fabricated thin sample cells ($150\, \mu m$ height) with the
described substrates as the bottom plate and inserted a diluted
suspension of colloidal particles in a water-2,6-lutidine (WL)
mixture at critical composition, i.e. a lutidine mass fraction of
$c{^c}{_L} \approx 0.286$. Such mixtures have a lower critical point
at $T_C\approx 307\, K$ \cite{bey85}. As colloids we used negatively
charged melamine spheres (MF) with radius $R=1.35 \mu m$
\footnote{MF-COOH-S1285, $R=1.35 \pm 0.05\,  \mu m $ microparticles
GmbH, Berlin, Germany. According to the manufacturer the surface
potential in water is 70-100\,  mV. }. Due to their high surface
charge density the particles are strongly hydrophilic, i.e. they
show a preference for water adsorption.

Interaction potentials between a single colloid and a substrate were
measured with total internal reflection microscopy (TIRM). The
entire sample cell was mounted onto a glass prism such that an
incident p-polarized laser beam ($\lambda =473\, nm$, $P\approx2\,
mW$) is totally reflected at the substrate-fluid interface. Under
these conditions an evanescent field is created which exponentially
decays into the fluid. Our experiments were performed with a
penetration depth of $153\, nm$ by adjusting the angle of incidence
accordingly. When the height $z$ of the colloid above the surface is
in the region illuminated by the evanescent field, it will partially
scatter the evanescent light. For the chosen conditions, evanescent
light scattering on critical fluctuations in the mixture can be
neglected compared to the light scattered by the colloidal particle.
From the scattered intensity, which is monitored with a
photomultiplier, the particle-substrate height distribution $P(z)$
can be inferred. Employing the Boltzmann factor, the height-resolved
interaction potential for a colloid close to the substrate is
derived. The lateral motion of the particle was reduced to about $
\pm 1\, \mu m$ with a weakly focussed laser beam ($\lambda=532\,
nm$) acting as an optical tweezers from above. Since this value is
orders of magnitude smaller than the lateral extension of the
chemical gradient, the boundary conditions can be considered as
homogeneous on the area probed during a single measurement. For
further details regarding TIRM and the experimental setup we refer
to the literature \cite{her08,wal97,pri90}.

Temperature control of the sample cell was achieved by a two-step
procedure. We connected the sample with a copper frame to a heat
bath operated at a constant temperature slightly below $T_C$. In
addition, we used an electrical heater, which was connected to a
temperature controller. In contrast to previous experiments, where
the temperature of the binary liquid mixture was stabilized with
respect to a platinum resistor placed outside the liquid, here the
light scattering intensity from the critical fluctuations was used
as input for the temperature-controller \cite{sch04}. For this
purpose an additional laser beam ($\lambda =658\, nm$) was coupled
into the cell to propagate parallel to the substrate in the fluid.
With this setup we achieved a temperature stability of about $\pm
2\, mK$. Since the scattering signal tends to diverge at the
critical temperature, we can determine $T_C$ with a significantly
improved accuracy of $\pm 5\, mK$.

\begin{figure}
\includegraphics[width=8.5cm]{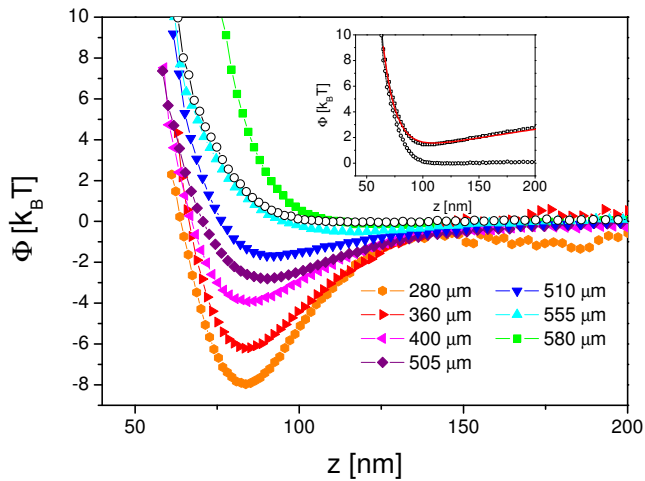}
\caption{(Color online) Interaction potentials between a MF particle
and a silica substrate with a chemical gradient, obtained at
different lateral positions $\Delta x$. Closed symbols were taken at
$T_{C}-T=220\, mK$ corresponding to $\xi=19\, nm$. The open symbols
show a measurement for $T_{C}-T=1.50\, K$ where critical Casimir
forces are negligible. Inset: interaction potential as measured
(squares) and after subtraction of the linear contributions
(circles) which are due to optical and gravitational forces. The
solid line is a fit to Eq.~\ref{eq:dlvo} with $A=2770\, k_BT$,
$\kappa^{-1}=11.1\, nm$ and $G^*= 13.5\, k_BT/ \mu m$. \label{fig:2}
}
\end{figure}

The inset of Fig.~\ref{fig:2} (upper curve) shows a typical
interaction potential between a single MF particle and a surface far
below $T_C$ where critical Casimir forces are negligible. The shape
of the potential can be fitted to
 \begin{equation}\label{eq:dlvo}
\Phi (z)=A \exp(-\kappa z)+G^*z
 \end{equation}
with $A$ the amplitude of electrostatic interactions between the
negatively charged particle and substrate, $\kappa$ the inverse
Debye screening length of the mixture and $G^*$ the effective weight
of the colloid due to gravity and light pressure from the optical
tweezers. Since the linear contribution from gravitational and
optical forces does not vary between individual measurements, in the
following it has been subtracted from all data in this paper. Since
the particle-wall interaction potential (Fig.~\ref{fig:2} inset) is
well fitted by Eq.~\ref{eq:dlvo} and parameters in agreement with
literature values \cite{gal92,gru01}, possible contributions from
van der Waals forces are negligible in the present experiment. A
more detailed discussion can be found in \cite{gam09c, dan07}.

Interaction potentials at constant temperature $T_{C}-T=220\, mK$
and different lateral positions $\Delta x=x-x_0$ relative to the
substrate (with $x_0$ a reference position at the strongly
hydrophilic site of the gradient) are shown in Fig.~\ref{fig:2}.
Since the colloidal particle is strongly hydrophilic, symmetric BC
should apply at small values of $\Delta x$, i.e. on the strongly
hydrophilic side. Under these conditions critical Casimir forces are
attractive. In combination with the short-ranged electrostatic
force, this leads to potential wells with depths of several times
the thermal energy $k_{B}T$. With increasing $\Delta x$, i.e. upon
approaching the hydrophobic side of the gradient, the BC become
increasingly asymmetric and the critical Casimir forces become
weaker. Accordingly, the potential wells become shallower and are
shifted towards larger distances. Close to the hydrophobic region of
the gradient lutidine is preferred by the substrate and the critical
Casimir forces should be repulsive. Indeed for $\Delta x=580\, \mu
m$, such a repulsion is observed in our data, as can be seen by
direct comparison with the particle-wall interaction potential far
below $T_C$ (open symbols) where critical Casimir forces are
negligible.

\begin{figure}
\includegraphics[width=8.5cm]{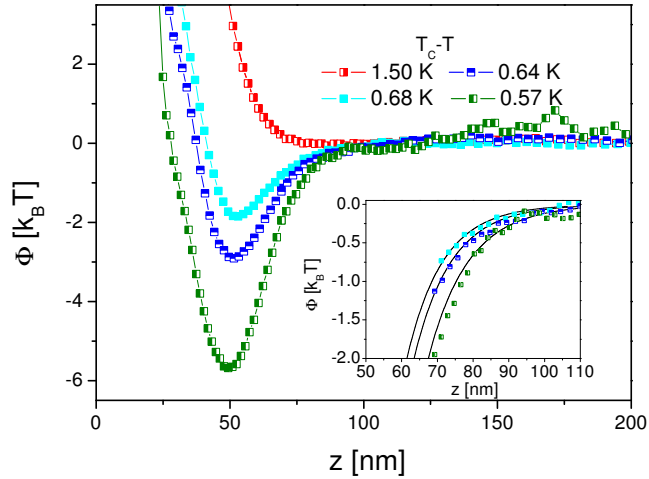}
\caption{(Color online) Temperature dependence of critical Casimir
forces at fixed position $\Delta x=0$. The inset shows the distance
range where electrostatic interactions are negligible. Fits to
theoretical predictions for $(- -)$ BC are shown as solid lines.}
\label{fig:3}
\end{figure}

For $(- -)$ and $(- +)$ BC the critical Casimir potential of a
colloidal sphere with radius $R$ at height $z$ above a homogeneous
surface is given by \cite{her08, han98}
\begin{equation}
 \Phi_{Cas}\left(z,T\right)=\frac{R}{z}\vartheta\left(\frac{z}{\xi}\right)
\label{eq:1}
 \end{equation}
with the correlation length
\begin{equation}
\xi =\xi_0 \left ( \frac{T_C-T}{T_C} \right )^{-\nu},   \label{eq:2}
\end{equation}
$\xi_0$ reflecting the typical length scale set by the
intermolecular pair potential in the mixture, $\nu = 0.63$ the
critical exponent of the 3D Ising universality class and $\vartheta$
the corresponding scaling functions which have been inferred from
Monte-Carlo simulations \cite{vas07} \footnote{The Derjaguin
approximation was used to adapt simulation results for wall-wall
geometry to the sphere-wall geometry of the experiment. This is
justified since R is much larger than its distance $z$ and the
maximum correlation length $\xi_{max}=40nm$.}. To confirm, that the
potentials indeed result from critical Casimir forces, we first
investigated the temperature-dependence of the potential at ($\Delta
x=0$) where $(- -)$ BC apply (Fig.~\ref{fig:3}). In the inset we
show the experimental data for the region where electrostatic and
van-der-Waals interactions are negligible. As solid lines we plotted
the fits according to Eq.~\ref{eq:1} which show good agreement. It
should be emphasized that only $\xi_{0}$ has been used as an
adjustable parameter. Best agreement with our experimental data was
found for $\xi_{0}\approx 0.2\, nm$, which is in good agreement with
other measurements in critical water-lutidine mixtures
\cite{gam09c,gul72}.

\begin{figure}
\includegraphics[width=8.5cm]{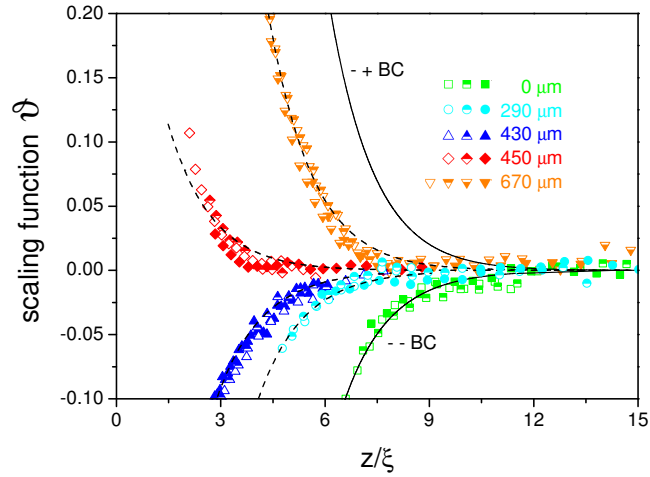}
\caption{(Color online) Measured scaling functions for different
lateral positions $\Delta x$ on the substrate. Symbols of the same
kind but with different fillings correspond to measurements at
different temperatures and collapse to a single curve each. The
square symbols were determined from the corresponding curves in
Fig.~\ref{fig:3}. Theoretical calculations for the limiting cases of
$(- -)$ and  $(- +)$ BC are shown as solid black lines. \cite{her08,
vas07}. Dashed lines represent the same curves shifted along the
$\frac{z}{\xi}$-axis to obtain best agreement with data for weaker
adsorption preference.} \label{fig:4}
\end{figure}

According to Eq.~\ref{eq:1} the information about the BC is entirely
encoded in the scaling function $\vartheta$. Therefore, we
determined $\vartheta$ from the measured critical Casimir
interaction potentials for different substrate positions $\Delta x$
(Fig.~\ref{fig:4}). Note that symbols with identical shape but
different fillings correspond to scaling functions obtained at the
same position $\Delta x$ but for different temperatures. Data taken
at different temperatures collapse onto a single curve in this
representation. With increasing $\Delta x$ the scaling functions
change systematically from negative to positive values. This is
consistent with the sign change of critical Casimir interactions
observed along the chemical gradient as shown in Fig.~\ref{fig:2}.
For comparison we added as solid lines the theoretical predictions
for the scaling functions for $(- -)$ and $(- +)$ BC. As can be
seen, the measured values for $\vartheta$ lie in between these
limiting cases. On the hydrophilic side of the substrate we
obviously reached $(- -)$ BC while we did not reach $(- +)$ BC on
the hydrophobic side. This indicates that the lutidine adsorption on
the OTS treated substrate is not saturated.

Scaling behavior is observed for all positions $\Delta x$ which is
{\it not a priori} clear because Eq.~\ref{eq:1} is strictly valid
only for $(- -)$ and $(- +)$ BC \cite{sch08}. This indicates that
additional scaling variables which may arise in the presence of
undersaturated adsorption layers are not relevant for data collapse
at the $\frac{z}{\xi}$-range sampled in our experiments
\cite{cho02}. Mean field theory calculations predict that the
scaling functions for BC close to the strong adsorption limit can be
obtained by a shift along the $\frac{z}{\xi}$-axis
\cite{mac09,bin83}. This is in remarkable agreement with the dashed
lines in Fig.~\ref{fig:4} which just correspond to shifted
$\vartheta$ functions for $(- -)$ and $(- +)$ BC obtained by a least
mean square fit to the data.

At present, it remains unclear how experimentally accessible
parameters for the quantitative characterization of boundary
conditions can be related to e.g. the surface field $h_1$ which is
often used to theoretically describe continuously varying BC
\cite{mac03,des95,dur87}. Ellipsometry studies on critical
adsorption of binary mixtures under weak surface field conditions
suggest that $h_1$ is proportional to the surface energy difference
of the two liquid components \cite{cho02}, while other approaches
tried to connect the surface field with the difference in the
chemical potential \cite{des95}. We hope that our work will
stimulate further theoretical investigations in this direction.

In summary, we have shown that critical Casimir forces can be
continuously varied by appropriate BC of the confining surfaces.
Experimentally, this was achieved by lateral variation of the
surface coverage of a single layer of OTS molecules on the substrate
which leads both to a change of the magnitude and the sign of
critical Casimir forces between a colloidal particle and the
surface. In addition to the exquisite temperature dependence, this
remarkable sensitivity on the surface properties of the interacting
objects distinguishes critical Casimir forces as a versatile
interaction type which adds novel perspectives to the use of
colloidal suspensions as model systems but also opens new
possibilities for the fabrication of colloidal crystals which hold
significant interest for technical applications.

\acknowledgments We thank S. Dietrich, A. Macio\l ek, T. Mohry, and
A. Gambassi for stimulating discussions, T. Geldhauser for
assistance with AFM measurements and the Deutsche
Forschungsgemeinschaft for financial support.


\end{document}